# Sampling of conformational ensemble for virtual screening using molecular dynamics simulations and normal mode analysis

G. Moroy [1,2,#], O. Sperandio [1,2,#], S. Rielland [1,2], S. Khemka [3], K. Druart [1,2], D. Goyal [3], D. Perahia [3], M. A. Miteva [1,2,*]

Gautier Moroy[1,2,#], Olivier Sperandio[1,2,#], Shakti Rielland[1,2], Saurabh Khemka[3], Karen Druart[1,2], Divij Goyal[3], David Perahia[3] & Maria A Miteva*

[1]Université Paris Diderot, Sorbonne Paris Cité, Molécules Thérapeutiques In Silico, INSERM UMR-S 973, Paris, France

[2]INSERM, U973, Paris, France

[3]Laboratoire de Biologie et de Pharmacologie Appliquée (LBPA), CNRS UMR 8113 ENS de Cachan, Cachan, France

* Author for correspondence:
phone: +331 57 27 83 92
fax: +331 57 27 83 22
email: maria.miteva@univ-paris-diderot.fr

# Authors contributed equally

**Abstract :**

**Aim:** Molecular dynamics simulations and normal mode analysis are well-established approaches to generate receptor conformational ensembles (RCEs) for ligand docking and virtual screening. Here, we report new fast molecular dynamics-based and normal mode analysis-based protocols combined with conformational pocket classifications to efficiently generate RCEs. **Materials & methods:** We assessed our protocols on two well-characterized protein targets showing local active site flexibility, dihydrofolate reductase and large collective movements, CDK2. The performance of the RCEs was validated by distinguishing known ligands of dihydrofolate reductase and CDK2 among a dataset of diverse chemical decoys. **Results & discussion:** Our results show that different simulation protocols can be efficient for generation of RCEs depending on different kind of protein flexibility.

**Graphical Abstract**

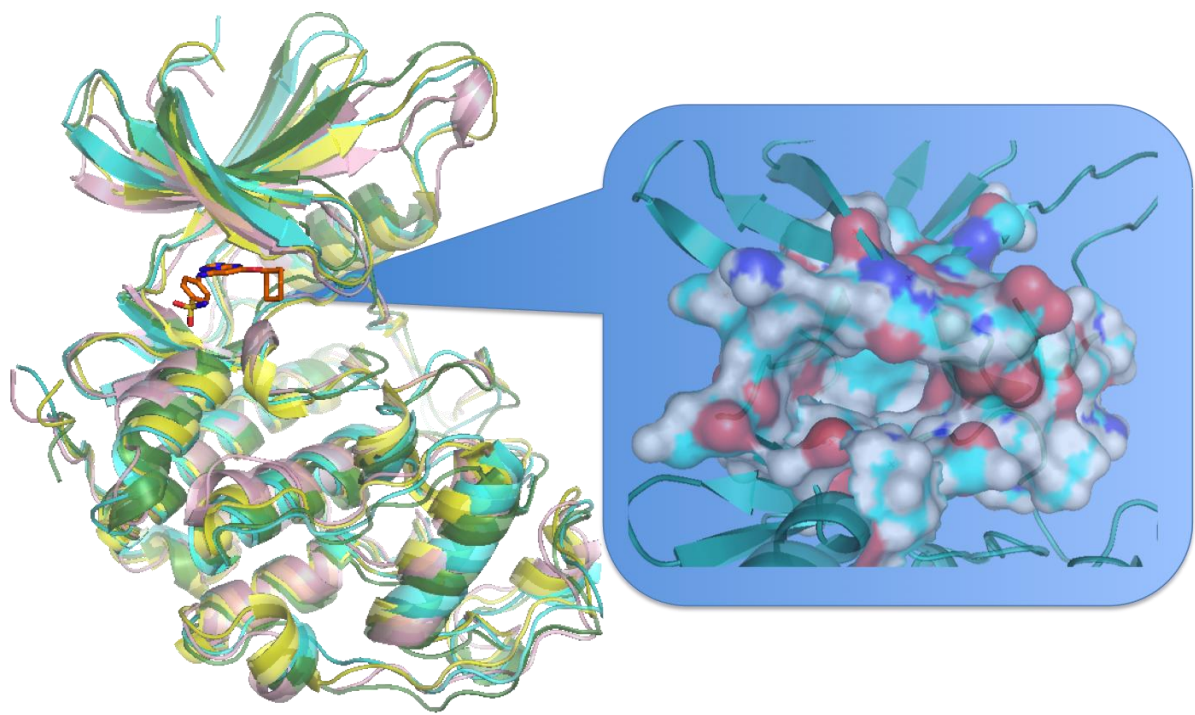

**Introduction**

Proteins are inherently flexible macromolecules and they undergo various conformational changes in a wide range of amplitudes and timescales to exert their functions. This is of particular importance for protein – ligand interactions [1,2]. The flexibility and conformational changes of both the ligand and the receptor occurring upon binding must be taken into account to correctly estimate the thermodynamics of the binding reaction [3]. Following the complexity to consider the conformational changes of bound and unbound states of the system, structure-based drug design studies often neglect these phenomena by using a static representation of the protein target. In order to overcome this important drawback protein flexibility has been integrated into ligand docking and structure-based virtual screening [4,5] in various ways, including side-chain flexibility [6,7] soft docking, induced fit [8,9] or conformational ensemble-based docking [10,11]. A correct incorporation of protein dynamics for drug design is still a difficult task. It has been shown that in many cases including protein flexibility may lead to higher rates of false positives as a large number of putative ligands can be accommodated into different conformations of the binding pocket [12-14].

Protein conformational ensembles are frequently used to include protein flexibility in current ligand docking and structure-based virtual screening approaches. It corresponds to our modern understanding that proteins exist in an ensemble of conformational substates [15,16]. In fact, extensive experimental and computational knowledge have been acquired [17,18] demonstrating that the protein conformational population shift scenario is regularly valid for ligand recognition [19,20] while induced fit is dominating in cases of extremely high ligand concentration [21]. These observations justify the widely used in silico approaches generating protein conformational ensembles prior to docking and allowing to probe small-ligands binding into different protein conformations individually [22,23]. However, the success of such exercise strongly depends on the quality of sampling. Experimental protein structures determined by X-ray crystallography or NMR can be used as receptor structures for docking or virtual screening (VS) [24,25 ,26-29]. In some cases, available experimental conformations may not be sufficient to represent various ligandable conformations of the binding site for correct prediction of accommodation of new ligands. Further, efficient generation of suitable ensemble of modeled receptor conformations to be used prior to docking is still challenging despite the various modeling techniques that have been employed [30,31].

Critical reviews on current approaches for generation of receptor conformational ensemble (RCE) have been reported [32,33]. Usually protein receptor conformations are identified using MD simulations [34-40]. McCammon and co-workers pioneered the Relaxed Complex Scheme (RCS), which is based on MD simulations to account for protein flexibility prior to ligand docking and VS [41,42]. Another MD-based approach, Limoc, aims at sampling RCE appropriate for accommodating ligands, which are chemically and structurally diverse and thus unbiased toward a particular class of ligands [43]. Recently we performed successful ligand profiling of drug metabolizing enzymes sulfotransferases by employing docking to RCE generated by MD simulations combined with hierarchical conformational clustering of different binding site conformations [44].

Another promising approach is to explore collective movements-based conformational changes [45,46]. Previously, we generated RCE via normal-mode analysis (NMA) and elaborated rules for the selection of several receptor conformations suitable for docking/VS by discarding the extremely altered binding site conformations while retaining diverse binding pockets [47]. Recently Leis and Zacharias studied receptor flexibility using an efficient potential grid representation of the receptor deformed in selected collective movements from NMA in order to consider global structural changes upon ligand binding [48]. In the same line, Abagyan and co-workers reported the original approach LiBERO relying on the use of ligand information for selecting the best performing RCE derived from NMA or Monte Carlo simulations [14]. However, taking into consideration a large number of modeled conformations may lead to less predictive VS results than the best performing crystal or NMR structures due to generation of non-native protein-ligand conformations [13,14,33,49-51]. Identifying the best performing RCEs is a complex task because usually frequently occurring protein conformations promote better binding conditions for different ligands while in some cases, rarely occurring protein conformation can be preferred for ligand binding [22].

Here, we focused on generating RCEs by using fast MD-based and NMA-based techniques combined with conformational pocket classifications. We assessed the performance of MD-based and NMA-based sampling dependent on two different kind of protein flexibility: local binding site flexibility and flexibility related to large collective protein motions. The two proteins studied here, DHFR and CDK2, are typical examples for such different flexibility of their binding sites. The druggability of the generated RCEs for the two proteins was analyzed and the performance of the RCEs was validated by distinguishing known ligands of DHFR and CDK2 among a dataset of diverse chemicals decoys. Several modeled conformations of DHFR and CDK2 were found to perform better than the used X-ray structures. The proposed protocols based on short MD simulations in implicit solvent (compared to other protocols based on long MD [52]) or based on NMA can be used for fast and efficient generation of RCEs that can be practical for structure-based virtual screening.

## Methods

**Target preparation**

*Dihydrofolate reductase (DHFR)*

The superposition of 40 X-ray human DHFR structures (the apo structure PDB ID: 1PDB and 39 ligand-bound structures) show that the main-chain conformation of the active site is conserved, and only local conformational changes including side-chain re-orientations upon ligand binding are present (e.g. R28, F31, K68, R70). The methotrexate (MTX)–bound structure PDB ID 1U72 was chosen (X-ray resolution 1.9 Å) for MD simulations and NMA. The structure 1U72 is very similar to the apo structure 1PDB (X-ray resolution 2.2 Å), the two structures show RMSD between all atoms of 0.58 Å. Different positions have been found for the residues F31 and R70 of the binding pocket. We selected the X-ray structure PDB ID 1U72 because of its best position of the key residue F31 for aromatic interactions (pi-pi or pi-T stacking) with bound ligands. The co-crystalized ligand MTX was removed for MD simulations and NMA.

*CDK2*

The superposition of 9 X-ray structures of human CDK2 co-crystalized with ligands and two apo X-ray structures (PDB ID: 2jgz and 1w98) (the resolution of these X-ray structures is below 2.9 Å) suggested that 2C6T represents an intermediate structure among several holo and apo structures of CDK2 ((Supplementary Figure S1), thus, it was appropriate as a starting point for MD simulations and NMA. The co-crystallized inhibitor triazolopyrimidine was removed for MD simulations and NMA.

Although the apo form of a protein is usually used to generate RCE when no ligand-bound conformation is available, it has been observed in our previous [44,53] and other studies [43] that virtual screening using RCE generated by MD simulations starting from a ligand-bound protein structure outperformed virtual screening to RCE generated by MD simulations on the apo-protein structure. In this work we generated RCE starting the MD simulations or NMA from a protein structure extracted from a ligand-bound structure. Hydrogen atoms were added to the two proteins using CHARMM v. c35b4 [54] following the protonation distribution predicted by the pKa calculations with the PROPKA program [55].

In order to calculate the pocket volumes of DHFR and CDK2 for the large number of protein conformations used for virtual screening, first we used the Protomol utility implemented in the software Surflex [56] (Version 2.1) to generate the binding pocket. The output file is then used to compute the volume (MSMS package implemented in Chimera [57]. We also analyzed the druggability of the best performing models and X-ray structures by calculating the pocket volume and Drug score using DoGSiteScorer [58]. DoGSite calculates several pockets descriptors and employs support vector machine method to return a score of druggability between 0 and 1 (0 – nondruggable, 1 - druggable).

**Molecular dynamics simulations**

MD simulations were performed following exactly the same protocol for DHFR and CDK2 using CHARMM v. c35b4 version [54]. We used the all atoms PARAM27 force field [59] with CMAP correction. The solvation was taken into account by the Generalized Born implicit solvent function FACTS [60]. We employed MD simulations with implicit solvent as it has been recently demonstrated its reliability for protein modeling [61].

Non-bonded interactions were truncated in a cut-off distance of 12 Å with a shift function for electrostatics and a switch function for the van der Waals interactions. The protein structures were initially minimized using 500 steps of steepest descent (SD) algorithm followed by 500 steps of conjugate gradient (CG) algorithm. Distances between heavy atoms and hydrogen atoms were constrained using the SHAKE algorithm [62] allowing a time step of 2 fs. The system was heated during 200 ps to reach 300 K and then equilibrated during 400 ps with a temperature window of 300±10 K. The production time was 4 ns for each MD simulation run. We have 4 independent trajectories per protein with different initial velocities.

**Normal mode analysis**

To generate protein structures along single individual normal mode the following procedure was applied for DHFR and CDK2. The hydrogen atoms were first built and the whole complex was energy minimized using the CHARMM program [54] and the force field PARAM27 [59], using successively SD, CG and Adopted Basis Newton-Raphson (ABNR) algorithms. To avoid important deviations from the crystal structure, harmonic constraints were applied to all atoms with a force constant that was progressively decreased from 250 to 0 kcal.mol$^{-1}$.Å$^{-2}$ during the SD minimization. Then the minimization was continued without constraints until an RMS energy gradient of 10$^{-5}$ kcal.mol$^{-1}$.Å$^{-2}$ was reached. The normal modes were computed using the DIMB method [63], as implemented in CHARMM. Electrostatic interactions were treated with a 4r-dependent dielectric constant and a short switching function (applied between 6.0 and 8.0 Å) to avoid the shrinkage of the protein. We analyzed the first 34 internal modes (from 7 to 40) since the lower frequency modes are usually the most responsible for important conformational changes [64-66].

The atoms of the initial conformation (the minimized X-ray structures of CDK2 and DHFR) were displaced along the first 34 internal lowest frequency eigenvectors in both directions by increments of 0.2 Å until reaching a mass-weighted root mean square deviation (MRMSD) of 2 Å (or -2 Å) with respect to the initial conformation using the VMOD module of CHARMM. To distinguish between the two directions of an eigenvector, positive and negative values of MRMSD are used. To obtain energetically relaxed structures for each displacement, short energy minimizations (100 steps of SD with harmonic constraints and followed by CG and ABNR until getting a gradient of 5.10$^{-1}$ kcal.mol$^{-1}$.Å$^{-2}$) were performed on the sampled structures. This ultimately yielded 714 structures (34x21), 21 per normal mode (in the two directions of an eigenvector) including the initial structure minimized in the same conditions. The procedure used is described in more detail in [67]. For DHFR we additionally considered random linear combinations of 7 modes (combined modes) by varying randomly the amplitude along the normal mode vectors using an in-house CHARMM script. The generated structures were energy minimized as for single mode displacements.

**Protein conformational clustering**

In order to select representative structures for RCE we have used two non-supervised classification approaches as implemented in the R software [68]. For each protein the RMSD matrix was calculated for all atoms of the binding site and of the cofactor in the case of DHFR. First, we clustered different conformations of the binding sites by applying

Hierarchical Ascendant Classification (Hclust) on the obtained RMSD matrix using the aggregative method Ward as implemented in the R software [68]. Next, the K-means classification algorithm was also employed. K-means clustering aims to partition $n$ observations into $k$ clusters in which each observation belongs to the cluster with the nearest mean. The number of groups was chosen by a consensus of the criteria: visualization of the obtained Hclust trees, the intra-group variance for the K-means procedure, the Dunn index estimating the density and separation of the groups, and the Davies Bouldin index estimating the group dispersion. Finally, we took the centroid structure of each cluster in order to define a representative set of protein conformations for subsequent virtual screening experiments.

**Virtual screening**

**Data set preparation**

Actives compounds on both DHFR and CDK2 were taken from the DUD data set [69] (version 2010). Only one tautomeric state was considered for each ligand. The selection of the unique tautomers was made using Chemaxon Marvin sketch (version 2010) (www.chemaxon.com) and the option major macrospecies at pH 7.4. The Chembridge DiversityChem (version 2010) was used to prepare the decoy data set starting with 50 000 compounds. All datasets were first filtered for drug-likeness using the software FAFDrugs2 [70] and standard physicochemical properties with the following ranges: 150 < Molecular Weight < 750, 0 < Number of Hydrogen Bond donors < 8, 0 < Number of Hydrogen bond acceptors < 12, 0 < number of rotatable bonds < 13, 0 < polar surface area < 160, -5 < XlogP3 < 6. Furthermore a chemical diversity criterion was used to ensure the chemical representativeness of the chosen chemical structures and prevent the overrepresentation of some given chemical series. Such chemical diversity was made possible using a combination of the program Cactus (http://cactus.nci.nih.gov) and Subset (http://cactus.nci.nih.gov) that were used to respectively calculate chemical fingerprints of the compounds and ensure a maximal chemical similarities between them based on a Tanimoto distance equal to a maximum of 0.75. As a consequence, the obtained data sets were composed of 51 CDK2 inhibitors, 191 DHFR inhibitors, and 2739 decoys form the Chembridge subset, all diverse chemically and possessing acceptable physicochemical profiles. The resulting compounds were finally generated in 3D using Frog2 [71].

**Virtual screening experiments**

AutoDock Vina program was used to perform flexible ligand docking [72]. All atoms that do not belong to the proteins are removed except for the NADPH cofactor in the active site of DHFR. The protonation states of protein titratable groups were computed using PROPKA program [55]. Gasteiger charges were added to each atom using the AutoDockTools package. We used grid resolution of 1 Å, number of binding modes of 10 and exhaustiveness of 8. The search spaces have been centered on the binding sites of the proteins with cubic dimensions (26x26x26 Å for CDK2 and 20x20x20 Å for DHFR).

**Results & discussion**

We used human DHFR and CDK2 to evaluate the performance of the developed RCE generating protocols as model proteins since they are important therapeutic targets and they have been shown to be challenging among other targets when including flexibility *via* different protein conformations for virtual screening [43,73]. DHFR is an enzyme, which converts dihydrofolate into tetrahydrofolate and plays an essential role in cell metabolism and cellular growth. It has been validated as an anti-cancer target in a number of studies (see for ex. [74,75]). DHFR shows local flexibility of the active site as several side chains change their conformations depending on bound ligands (PDB ID: 1u71, 1u72, 1dlr, 1dls). CDK2 is also an important anti-cancer target involved in central cell cycle functions [76] by interacting with cyclins through the S phase and thus participating in the initiation and the progress of the DNA synthesis. Thus far CDK2 has been extensively investigated and a number of inhibitors have been discovered [77-79]. The superimposition of nine CDK2 structures (PDB ID : 3ti1, 3tiy, 4erw, 4ez3, 4acm, 2xmy, 2xnb, 2x1n, 2c6t) bound to various ligands and two apo CDK2 structures (PDB ID: 2jgz and 1w98) shows that significant induced fit of the ATP-binding site occurs upon ligand binding. The most important conformational changes occur on the G-loop (ILE10-VAL16) closing or opening the ATP-binding site. In addition, such movement is observed between the two apo-structures (more closed 2jgz and more open 1w98 conformations) suggesting that such collective motion occurs at this region even without ligand binding.

Figure 1 shows the computational procedure used to generate and validate the RCEs for DHFR and CDK2. RCEs generated by MD simulations or NMA were assessed based on their performance to distinguish active and diverse decoy compounds by docking using AutoDock Vina [72] and by calculating the enrichment at 1%, 5% and 10% of the screened chemical library (percent of actives recovered). We have chosen AutoDock Vina because of its good performance of binding affinity prediction and speed [80]. In addition Vina is not very sensible to errors in the protonation behavior for various ligands [72] that may occur when one screens a large number of compounds.

**Molecular Dynamics Simulations and Receptor Conformational Ensemble for DHFR**

We ran four MD simulations (noted as MD1, MD2, MD3 and MD4) for human DHFR with different initial velocities. The calculated root-mean-square deviations (RMSD) of backbone atoms for the entire protein against the average MD structure were < 2 Å for the four trajectories ensuring thus the reliability of the MD simulated DHFR structures. MD2 and MD4 have shown larger fluctuations with RMSD < 2 Å for MD2 and < 1.5 Å for MD4, against RMSD < 1 Å for MD1 and MD3, respectively. We took 4000 snapshots from each MD trajectory of the entire production run (one conformation every 1 ps) for further consideration, in total 16000 MD generated conformations for DHFR. Our analysis focuses mainly on the plasticity of the binding site observed during the MD simulations. The list of 29 protein residues of the binding site is given in Supplementary material. Figure 2A shows the conformational space of the 16000 generated DHFR structures following the structural differences of their binding sites. It is seen that the four MD trajectories take a specific place inside the total binding-site conformational space. Similarly, the four trajectories can be distinguished on the RMSD map (Fig. 2B).

In order to extract a suitable RCE among the 16000 MD structures for ligand docking/virtual screening with diverse binding-site conformations, we employed two classification strategies based on the matrix of RMSD for all atoms of the binding site and the co-factor NADPH: Hierarchical Ascendant Classification (Hclust) and K-means clustering. Twenty six structures were finally retained to be probed for virtual screening experiments, 13 centroids obtained from the Hclust classification and 13 centroids obtained from the K-means classification. The resulted RMSD between the centroids was > 0.4 Å for Hclust and > 0.7 Å for K-means.

**Virtual screening for RCE generated by MD simulations for DHFR**

VS experiments were performed using docking-scoring approach in order to identify the MD conformations of DHFR, which better discriminate known binders from putative decoys. We ran 26 VS for the RCE of DHFR generated by MD simulations. The best results obtained for the MD centroid conformations using the two classification methods and the X-ray structure (PDB ID 1U72) are shown in Table 1. The structure MD_11281 (from MD3) (Fig.3) obtained by Hclust and K-means achieves better enrichment results than the X-ray structure following the enrichment at 5%. The other conformations do not show better performance than the X-ray structure. Overall the centroids obtained by K-means achieve better enrichment results than those obtained using the Hclust classification. Yet, the both classifications found various conformations (except MD_11281) with different pocket volumes and RMSD. Hence running several short MD simulations seems to be a pertinent approach in order to cover larger conformational space of the DHFR active site.

We performed structural analysis of the binding sites of the MD generated conformations and of the X-ray structure (Table 1). The volumes of the binding sites of the best performing MD conformations vary from 507 to 791 Å$^3$. The best structures show volumes of the binding pocket quite similar or smaller to that of the X-ray structure. Overall MD2 and MD4 generated conformations have volumes of the active site pocket (up to 1600 Å$^3$) larger than those of MD1 and MD3. The best structures were found from MD1 and MD3 showing RMSD of the backbone atoms during the trajectories < 1 Å.

**Normal mode analysis and receptor conformational ensemble for DHFR**
We have previously shown that including all atoms in NMA can be critical for a quasi-exhaustive simulation of possible changes that may occur in the binding site [47]. Here we analyzed the first 30 modes (from 7 to 36) for DHFR since the lower-frequency modes are usually the most responsible for important conformational changes [64-66]. The atoms of the initial conformation (the minimized X-ray structure of DHFR, PDB ID 1U72) were displaced along the first 30 lowest frequency eigenvectors (apart those corresponding to the 6 global translations/rotations) in both directions by increment of 0.2 Å until reaching a Mass Weighted Root Mean Square Deviation (MRMSD) of 2Å (or -2Å) with respect to the initial conformation. To distinguish between the two directions of an eigenvector, positive and negative values of MRMSD are used. Twenty one conformations were generated per mode, this yielded finally 630 structures.

Following the observation that the best performing MD conformations for DHFR have volumes of the active site similar to the X-ray pocket volume, for the first NMA ensemble we have chosen structures having volumes of the binding pocket within a range of the volume of the X-ray binding pocket structure ± 30%. This resulted in 144 conformations that were used for VS following the same protocol as for the MD generated structures. No conformation was found to retrieve more known ligands of DHFR than the used X-ray structure at 5 % of the screened library (results shown in Supplementary material Table S1). In fact, the minimization performed before NMA displaced the key residue F34 preventing thus correct ligand binding in the active site. We should note that NMA may not be the best approach to explore conformational changes of binding sites showing only local conformational changes, e.g. side-chain movements observed in other studies [45].

In such cases a combination of different modes touching the binding site can be helpful to increase the conformational space generated by NMA. We probed such a strategy to generate additional DHFR conformations. We generated 2000 structures using linear combinations with random amplitudes of 7 modes (11, 16, 18, 20, 22, 24, and 35) that opened the active site. Then we performed clustering on the RMSD of the binding pocket residues to decrease the number of structures to screen using the same strategy as for the MD conformation classification. The two procedures (Hclust and K-means) and the consensus of the four criteria (visualization of the obtained Hclust trees, the intra-group variance for the K-means procedure, the Dunn index, and the Davies Bouldin index) resulted in 8 clusters. We took the 8 centroid conformations for DHFR. Again the VS experiments have not found any centroid conformation able to perform better than the used X-ray structure at 5 % of the screened library (results shown in Supplementary Table S2). To check the availability of conformations with correct position of F34 we calculated the distances between F34 and V8 of the binding site (results shown in Supplementary material Fig. S2) and we found a large number of NMA conformations showing a position of F34 preferable for ligand binding. Then, we decided to perform known ligand-driven analysis in order to find the best NMA conformations. For this purpose we docked the 191 DHFR actives into all 2000 NMA structures. The best conformations showing average binding energy ≤ -8.5 kcal/mol calculated by Vina (see Supplementary Fig. S3) were used for VS. The results for the five best preforming structures are shown in Table 2. The best NMA structure 1 achieves better enrichment results than the X-ray structure following the enrichment at 5%. Thus, a training process in which protein structures are selected on the basis of their performance to reproduce preferable binding affinities, as in our case of DHFR ligands, or to reproduce experimentally known binding modes [43] can be a useful approach to a rational selection of RCE for virtual screening purposes.

**Molecular dynamics simulations and receptor conformational ensemble for CDK2**

In order to validate the developed MD and NMA protocols we assessed their performance on CDK2. The superposition of several CDK2 structures co-crystallized with diverse ligands and two apo forms confirms that significant induced fit of the ATP-binding site occurs upon ligand binding. The two regions involved in conformational changes are the hinge region (E81-H84) and importantly the G-loop (I10-V16). For CDK2 we used exactly the same protocol to run four MD simulations with different initial velocities as for DHFR. Similarly, we took 4000 snapshots for CDK2 from each MD trajectory of the entire production run (one conformation every 1 ps) for further consideration, in total 16000 MD generated conformations. In order to eliminate structural redundancy and to extract a suitable RCE among the 16000 MD structures of CDK2 for ligand docking and virtual screening, we employed the same classification strategy as for DHFR based on the matrix of RMSD for all atoms of the binding site using Hclust and K-means (the list of the 24 residues of the ATP-binding site is given in Supplementary material). Twenty structures were thus retained to be probed for virtual screening experiments, 10 centroids obtained from the Hclust classification and 10 centroids obtained from the K-means classification.

**Virtual screening for RCE generated by MD simulations for CDK2**

VS experiments were performed for CDK2 using the same docking-scoring protocol as for DHFR. We ran 20 VS for the MD-generated RCE of CDK2. The best results for the MD centroid conformations obtained by the two classification methods and the X-ray structure ID PDB 2C6T are shown in Table 3. The two structures MD_6677 and MD_7889 obtained by K-means perform equally than the X-ray one at 1% of the ranked library, and better than the X-ray one at 5% of the ranked library. The centroid conformations of CDK2 extracted by the K-means procedure achieve better enrichment results than those obtained using the Hclust classification. Again, the two classification procedures found diverse conformations with different pocket volumes within a range the X-ray pocket volume $\pm$ 30%. These results confirm the appropriateness to run short MD trajectories in parallel in order to increase the conformational space of binding pockets of studied receptor for diverse ligand binding.

**Normal mode analysis and receptor conformational ensemble for CDK2**

We analyzed the first 34 modes (from 7 to 40) for CDK2 being the lowest-frequency modes. The atoms of the initial conformation (the minimized X-ray structure of CDK2, 2C6T) were displaced along the first 34 lowest frequency eigenvectors in both directions until reaching a MRMSD of 2Å (or -2Å) with respect to the initial conformation. To distinguish between the two directions of an eigenvector, positive and negative values of MRMSD were used. Twenty-one conformations were generated per mode, this yielded finally 714 structures. The best performing MD-generated conformations for CDK2 have volumes of the binding site similar to the X-ray one, thus, for the generation of the NMA-based RCE for CDK2 we have chosen structures having volumes of the binding pocket within a range of the volume of the X-ray binding pocket structure ±30%. We thus selected 159 conformations that were used for VS following the same VS protocol as for the MD generated structures. As can be seen in Table 4 the CDK2 structure generated by the mode "35 -1" was the best one modeled by NMA (Fig.4). Although the enrichment obtained at 1% of the ranked library is better for the X-ray structure than that for the "35 -1" structure, the enrichment at 5% is increased twice. The CDK2 structure generated by the mode "24 0.8" performed similarly as the X-ray one. It is seen that in the case of CDK2, in contrary to DHFR, a simple NMA protocol combined with pocket-volume-based filtering was sufficient to find two new conformations with diverse binding site conformations yet with similar volumes. Such results can be expected taking into consideration the well-known collective movement of the G-loop covering the ATP-binding site of CDK2 supported by the large number of X-ray structures of CDK2 co-crystallized with different ligands [81] [82] as well as by previous NMA studies performed on CDK2 [47,83]. In fact, the ATP-binding site is located at the interface of two subdomains, thus, CDK2 constitutes a very appropriate case to use NMA for RCE generation, permitting to explore domain (subdomain) movements. The best performing CDK2 conformations suggest a movement of the G-loop (I10-V17).

In order to take into account the anharmonic effects arising for relatively large structural changes in NMA, an exploration of the energy surface along a given normal mode direction, or combination of linear modes, is necessary. For both proteins this was achieved by energy minimizations for successive displacements using an umbrella potential targeting a desired location. In our previous studies we have shown the reliability of generated structures by using this approach [47,67]. However, it is to note that considering only the lower-frequency modes for DHFR, showing only local active site flexibility, was not sufficient to generate active site conformations suitable for ligand docking, the used X-ray structure was better performing. In order to increase the conformational space of the DHFR binding site we used an additional combination of different modes touching the binding site, which resulted in the found NMA_1 conformation better performing than the X-ray one at 5% of the ranked library. In the case of CDK2, showing collective movement of the G-loop opening and closing the ATP-binding site, simple NMA was sufficient to generate binding site conformations well performing for ligand docking. In perspectives, the newly developed hybrid approach MDeNM (Molecular Dynamics with excited Normal Modes) combining NM and MD simulations [84] may overcome some of the above described limitations. MDeNM is based on kinetic excitation of collective motions described by a set of normal mode vectors within a standard MD simulation, thus coupling efficiently global and local motions.

**Druggability assessment of the generated RCE**

Previous studies have shown variable observations on correlations between identified protein conformations best performing for virtual screening and druggability of their binding pockets [44,47,73,85]. In fact, various strategies have been used to select the best RCE (by RMSD from starting structure, binding site volume, radius of gyration, cognate ligand size, flexibility descriptors among others), however, no method for selecting the best RCE was found similar to other studies [24]. Here we analyzed the performance of the generated RCE for virtual screening *vs* the druggability and volume of the binding pockets as computed by DoGSiteScorer [58] for DHFR and CDK2. The obtained results (Fig. 5) do not suggest a clear correlation between the calculated Drug scores and volume of the pockets and the best-obtained enrichments of actives retrieved at 5 % of the ranked chemical library. Clearly, criteria based on physicochemical and topological properties of the binding pockets as volume, polarity, shape, lipophilicity, presence of hot spots (e.g. key residues) and an overall druggability evaluation are critical for initial RCE selection. High enrichment (e.g. > 15 ) at 5% of the screened library was achieved here when using conformations with pocket volume between 500 and 800 $\text{Å}^3$ for DHFR and between 650 and 1000 $\text{Å}^3$ for CDK2, respectively. Further, the calculated Drug score 0.45 was predicted to be sufficient for DHFR and 0.7 for CDK2, respectively, to achieve better enrichment than 15. Following these results, it seems that the correlation between predicted druggability and docking/VS performance is target-dependent. Enrichment result can also depend on the used chemical library. In our study we probed the same diverse decoys for the two proteins while using large datasets of diverse actives for DHFR and CDK2 taken from DUD (see method section).

Recently, we have observed [44] that despite of very high druggability score of some holo X-ray structures, the obtained enrichments are not always satisfactory. The druggability score is a useful evaluation but it might be not sufficient for a final selection of the best receptor conformations. Keeping in mind that druggability assessment is target-dependent, additional criteria can be employed when there is available information for known ligands, structural data or biological activities. In this study, such a strategy helped to identify the best conformations of DHFR for virtual screening among the generated RCE by NMA. The position of F34 was critical to identify the most appropriate binding site conformations for docking and virtual screening. In the same line, it has been recently proposed that receptors found by using automatic iteration of the sampling-selection with Ligand-guided Backbone Ensemble Receptor Optimization (the ALiBERO method) are able to better discriminate active ligands from inactives in flexible-ligand VS docking experiments [14]. Thus, knowledge for active/inactive ligands can be very helpful for the selection of the most appropriate ensemble conformations for VS.

## Conclusion

We focused on generating RCEs by using fast MD-based and NMA-based simulations combined with two different conformational pocket classifications. For DHFR and CDK2 RCEs obtained by the K-means classification better discriminated known binders than those obtained using the Hclust classification. Our results confirmed the appropriateness to run short MD with implicit solvent in order to generate binding site conformations suitable for ligand docking and VS. Considering only the lower-frequency modes for DHFR, which shows local active site flexibility, was not sufficient to generate active site conformations suitable for ligand docking. In the case of CDK2, showing a collective movement of the G-loop close to the ATP-binding site, simple NMA successfully generated binding site conformations well performing for ligand docking and VS. These results suggest that for local flexibility short MD simulations are sufficient to explore the flexibility of the binding site for subsequent ligand docking and NMA can be more appropriate for protein targets expected to have collective motions involving the binding pocket.

## Future Perspective

Our study suggests that short MD simulations with implicit solvent are sufficient to explore local flexibility of protein binding site for ligand docking and VS. NMA can be more appropriate for protein targets expected to have collective motions involving the binding pocket. In perspectives, coupling efficiently global and local motions by hybrid approaches of MD and NMA may help to overcome some of the current limitations of RCEs.

**Acknowledgments**

We thank the INSERM institute, University Paris Diderot, CNRS and ENS de Cachan. SK and DG were financially supported by ARCUS program of Ministry of Foreign Affairs of France.

# Figures

**Figure 1. Computational procedure used to generate and validate the RCE for DHFR and CDK2.** MD: Molecular dynamics; NMA: Normal-mode analysis.

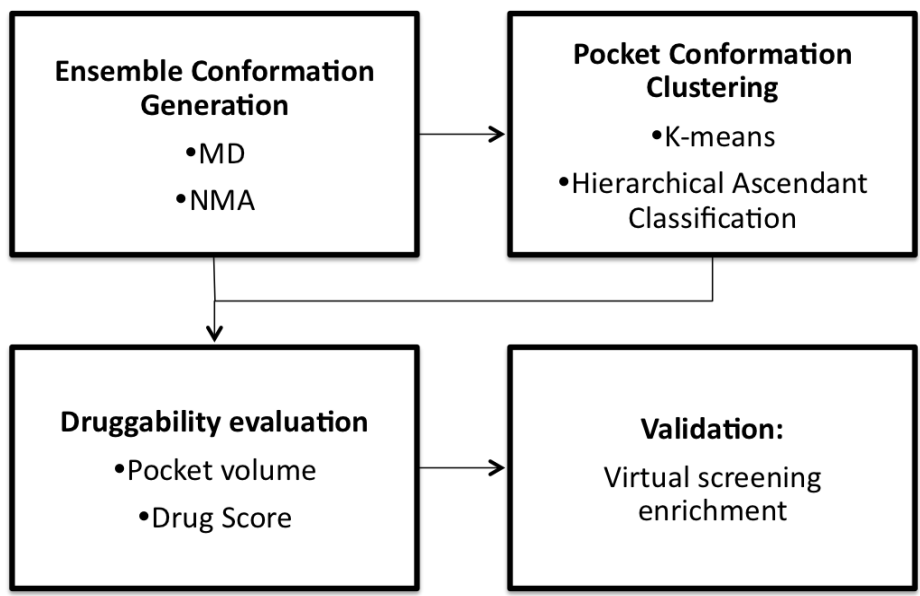

**Figure 2. Conformational space of DHFR explored by the MD simulations.** A. MultiDimensional Scaling representation of the conformational space of the 16000 MD generated structures for DHFR following their active site structural differences. MD1 in black, MD2 in red, MD3 in green, MD4 in blue. The location of the 13 centroids obtained by using K-means classification are highlighted in yellow. B. RMSD map for all active site atoms & co-factor NADPH between the 16000 MD structures. RMSD values color is progressing from red (0 Å for the same MD structure) to yellow; MD1 structure numbers: from 1 to 4000; MD2 structure numbers: from 4001 to 8000; MD3 structure numbers: 8001-12000; MD4 structure numbers: 12001-16000

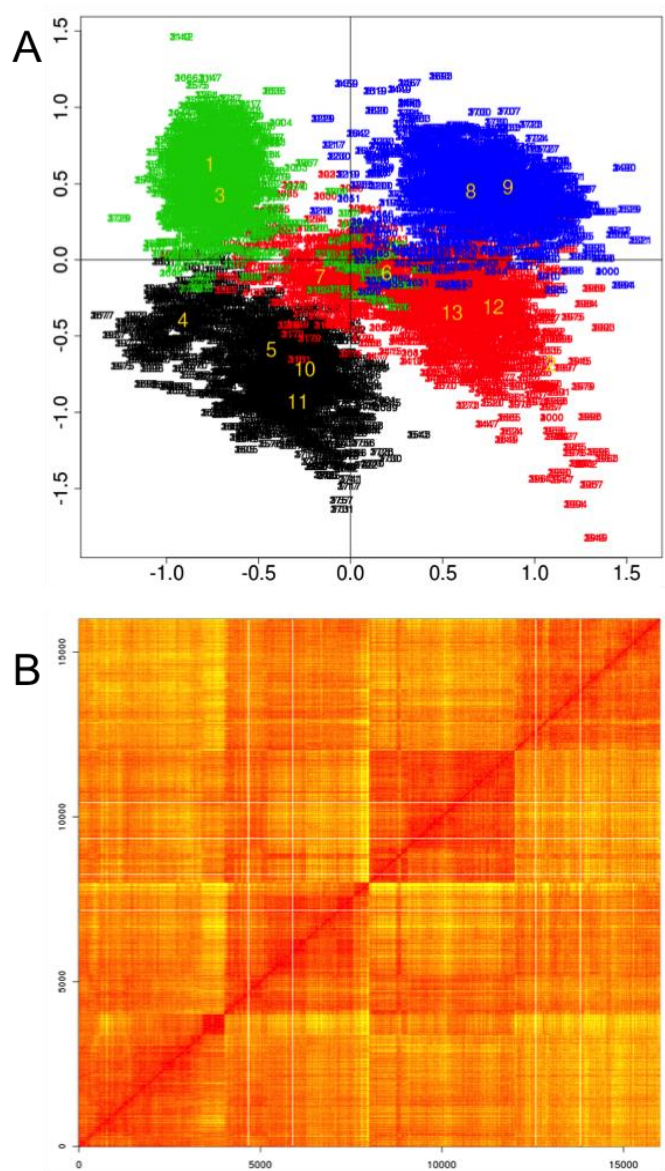

**Figure 3. Superimposition of human DHFR crystal structure PDB ID 1U72 (in yellow cartoon) and best performing structures of the generated RCE: MD_11281 (in cyan cartoon) and NMA_1 (in light violet cartoon).** The co-crystallized cofactor NADPH and methotrexate are shown in sticks colored in orange atom type. F34 is shown in sticks.

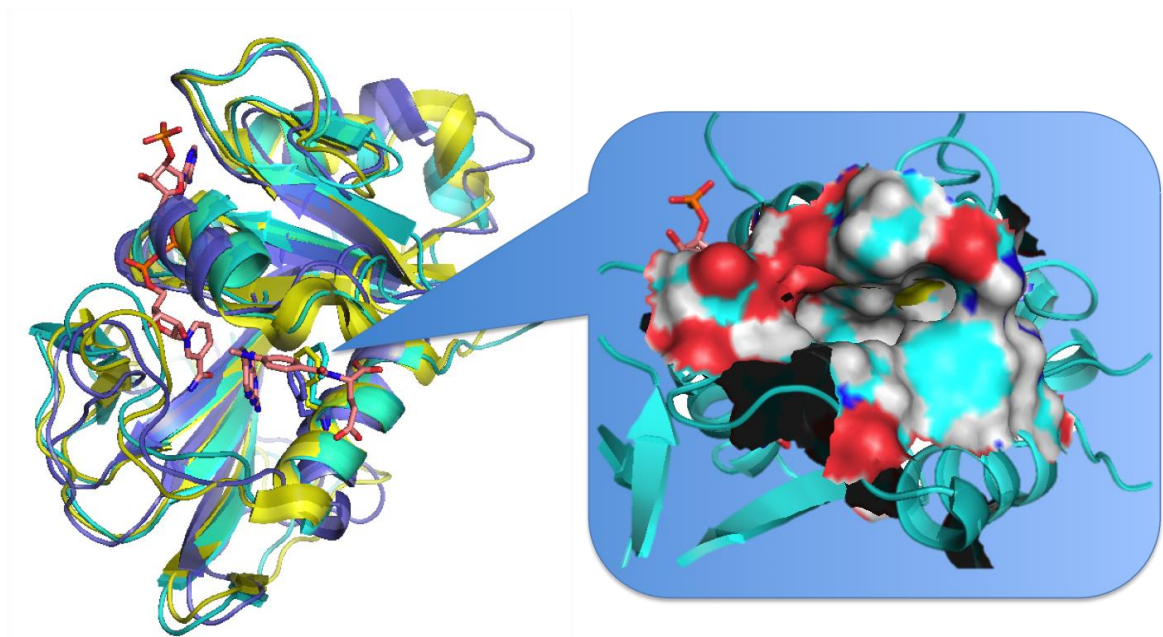

**Figure 4. Superimposition of human CDK2 crystal structure PDB ID 2C6T (in yellow cartoon) and best performing structures of the generated RCE for enrichment at 5%: MD_14709 (in cyan cartoon), MD_4749 (in green cartoon) and NMA_35-1 (in light pink cartoon).** The co-crystallized ligand triazolopyrimidine is shown in sticks colored in orange

atom type.

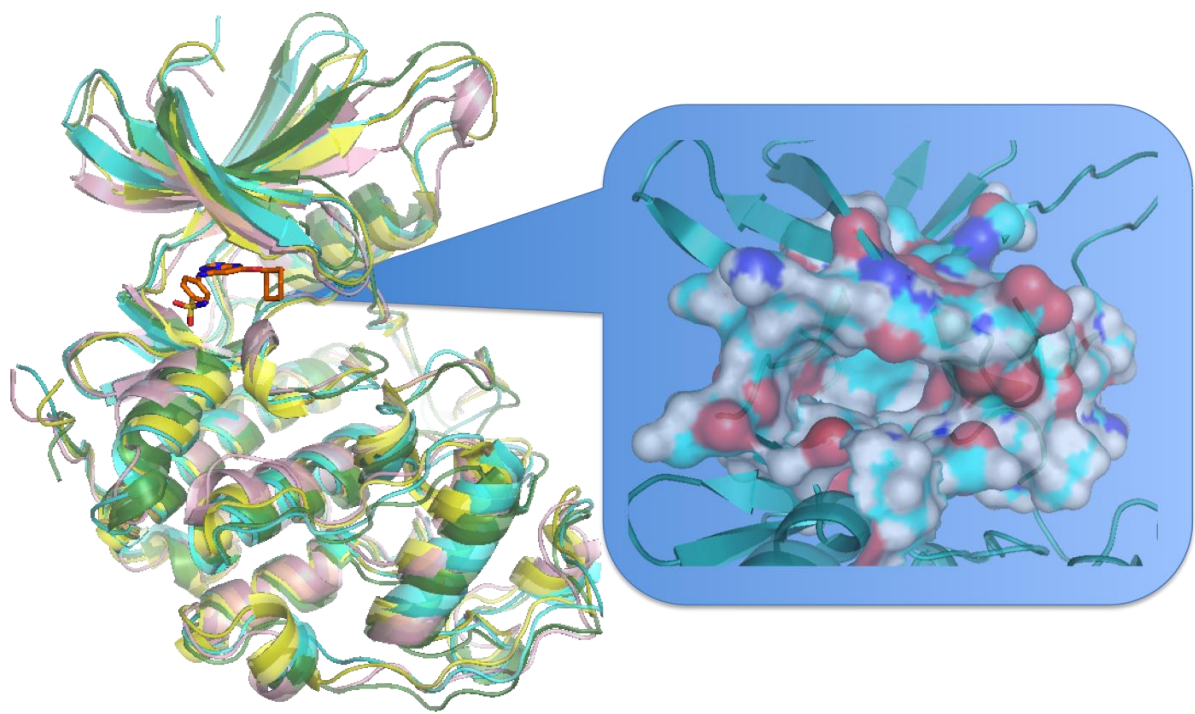

**Figure 5. Performance of the generated RCE *vs* druggability and volume of the binding pockets.** Volume values (in $Å^3$) are shown as red diamonds. Drug scores are shown as orange squares. A. for DHFR; B. for CDK2.

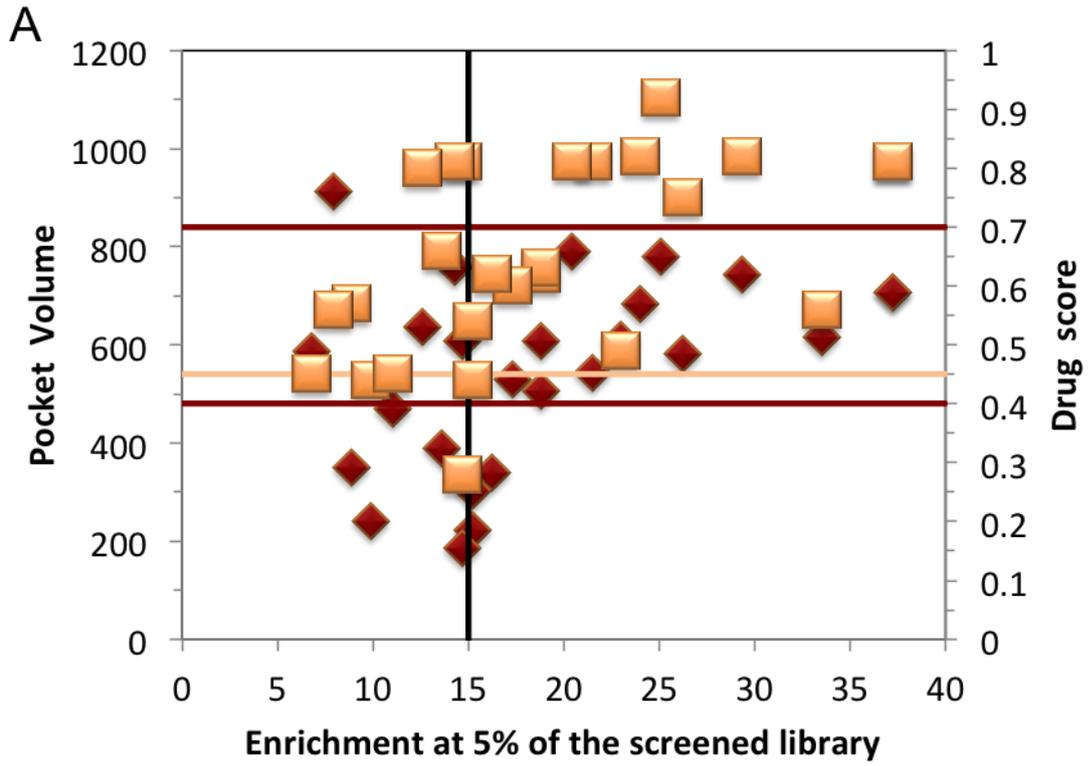

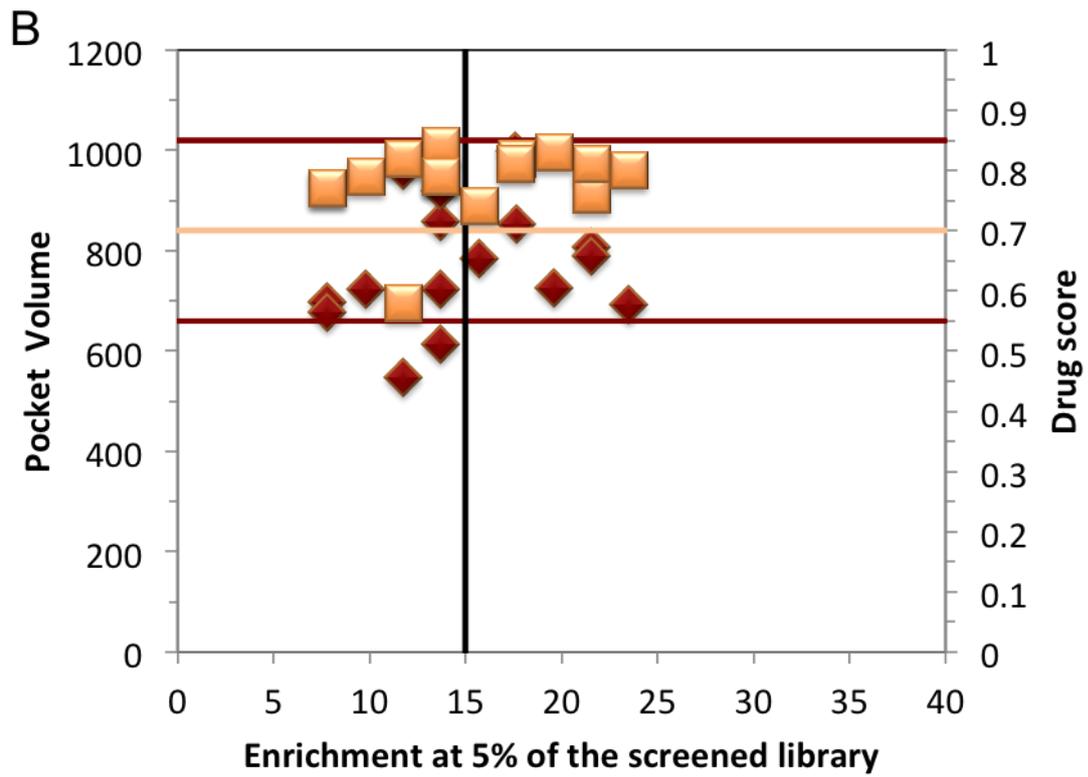

**Tables**

**Table 1.** Enrichment obtained on the X-ray and five best performing MD structures of DHFR for each classification at 1%, 5% and 10% of the screened library. Volume and Drug score values of the active sites are calculated using DoGSite webserver [58].

| Structure | 1% | 5% | 10% | RMSD of the binding site with 1U72 (Å) | Volume | Drug Score |
|---|---|---|---|---|---|---|
| **X-ray 1U72** | **8.4** | **29.3** | **55.0** | x | **744.4** | **0.82** |
| | | | | | | |
| **MD Hclust** | | | | | | |
| **MD_11281** | **9.9** | **37.2** | **56.5** | **1.79** | **708.2** | **0.81** |
| MD_3697 | 5.8 | 23.0 | 45.5 | 2.27 | 611.9 | 0.49 |
| MD_3889 | 6.3 | 21.5 | 36.6 | 2.08 | 544.0 | 0.81 |
| MD_3524 | 4.2 | 20.4 | 38.7 | 2.10 | 791.8 | 0.81 |
| MD_11469 | 2.1 | 14.7 | 30.9 | 2.41 | 608.1 | 0.81 |
| | | | | | | |
| **MD K-means** | | | | | | |
| **MD_11281** | **9.9** | **37.2** | **56.5** | **1.79** | **708.2** | **0.81** |
| MD_3685 | 6.3 | 26.2 | 52.4 | 2.22 | 583.0 | 0.75 |
| MD_3021 | 3.1 | 24.0 | 41.2 | 2.46 | 682.8 | 0.82 |
| MD_3396 | 7.9 | 18.8 | 30.9 | 2.06 | 609.5 | 0.62 |
| MD_3603 | 9.4 | 18.8 | 27.2 | 1.87 | 507.1 | 0.63 |

**Table 2.** Enrichment obtained on the five best performing DHFR structures generated by combined normal modes at 1%, 5% and 10% of the screened library. Volume and Drug score values of the active sites are calculated using DoGSite webserver [58].

| Structure | 1% | 5% | 10% | RMSD of the binding site with 1U72 (Å) | Volume | Drug Score |
|---|---|---|---|---|---|---|
| **NMA_1** | **8.9** | **33.5** | **66.5** | **2.54** | **617.4** | **0.56** |
| NMA_2 | 4.7 | 25.1 | 53.4 | 2.58 | 781.5 | 0.92 |
| NMA_3 | 3.7 | 17.3 | 26.7 | 2.63 | 530.4 | 0.60 |
| NMA_4 | 2.1 | 14.3 | 28.8 | 2.61 | 758.5 | 0.81 |
| NMA_5 | 2.6 | 12.6 | 16.2 | 2.75 | 637.8 | 0.80 |

**Table 3.** Enrichment obtained on the X-ray and five best performing MD structures of CDK2 for each classification at 1%, 5% and 10% of the screened library. Volume and Drug score values of the binding sites are calculated using DoGSite webserver [58].

| Structure | 1% | 5% | 10% | RMSD of the binding site with 2C6T (Å) | Volume | Drug Score |
|---|---|---|---|---|---|---|
| **X-ray 2C6T** | **5.9** | **13.7** | **35.3** | **x** | **858.4** | **0.8** |
| | | | | | | |
| **MD Hclust** | | | | | | |
| **MD_14709** | **2** | **21.6** | **31.4** | **1.05** | **808.5** | **0.76** |
| MD_8393 | 2 | 13.7 | 29.4 | 1.09 | 920.1 | 0.84 |
| MD_5877 | 2 | 13.7 | 23.5 | 1.02 | 613.4 | 0.79 |
| MD_4277 | 3.9 | 11.8 | 17.6 | 0.92 | 959.2 | 0.82 |
| MD_13157 | 2 | 11.8 | 25.5 | 0.99 | 548.4 | 0.58 |
| | | | | | | |
| **MD K-means** | | | | | | |
| **MD_4749** | **2** | **21.6** | **35.3** | **0.99** | **788.4** | **0.81** |
| **MD_13065** | **3.9** | **19.6** | **33.3** | **1.09** | **726.4** | **0.83** |
| **MD_15569** | **3.9** | **17.7** | **31.4** | **1.24** | **853.1** | **0.82** |
| **MD_6677** | **5.9** | **17.6** | **33.3** | **1.10** | **998.4** | **0.81** |
| **MD_7889** | **5.9** | **15.7** | **29.4** | **1.08** | **784.2** | **0.74** |

**Table 4.** Enrichment obtained on the five best performing NMA structures of CDK2 having volumes of the binding site within a range of the volume of the X-ray binding site ±30% at 1%, 5% and 10% of the screened chemical library. Volume and Drug score values of the binding sites are calculated using DoGSite webserver [58].

| Structure | 1% | 5% | 10% | RMSD of the binding site with 2C6T (Å) | Volume | Drug Score |
|---|---|---|---|---|---|---|
| **35 -1** | **2** | **23.5** | **37.3** | **1.25** | **693.3** | **0.80** |
| 24 0.8 | 3.9 | 13.7 | 31.4 | 0.88 | 723.7 | 0.79 |
| 23 -1.4 | 3.9 | 9.8 | 35.3 | 0.87 | 723.1 | 0.79 |
| 23 -1.6 | 3.9 | 7.8 | 31.4 | 0.92 | 698.1 | 0.77 |
| 22 0.6 | 2 | 7.8 | 33.3 | 0.83 | 678.2 | 0.77 |

**Supplementary Material**

**Text S1.**

Residues of the active site of DHFR.

I7, V8, A9, D21, L22, W24, P25, L27, R28, E30, F31, R32, Y33, F34, Q35, M52, T56, S59, I60, P61, K63, N64, L67, K68, R70, V115, Y121, T136, I138

**Text S2.**

Residues of the ATP-binding site of CDK2.

E8, I10, G11, E12, G13, T14, G16, V18, K20, V30, A31, K33, E51, L55, V64, F80, E81, F82, L83, L134, K142, A144, D145, N132

**Figure S1.** Superimposition of nine CDK2 structures (PBD ID: 3ti1, 3tiy, 4erw, 4ez3, 4acm, 2xmy, 2xnb, 2x1n, 2c6t) bound to various ligands and two apo CDK2 structures (PDB ID: 2jgz and 1w98). The structure of CDK2 bound to triazolopyrimidine is shown in magenta (PDB ID 2C6T). The co-crystallized ligand triazolopyrimidine is shown in sticks colored in magenta atom type.

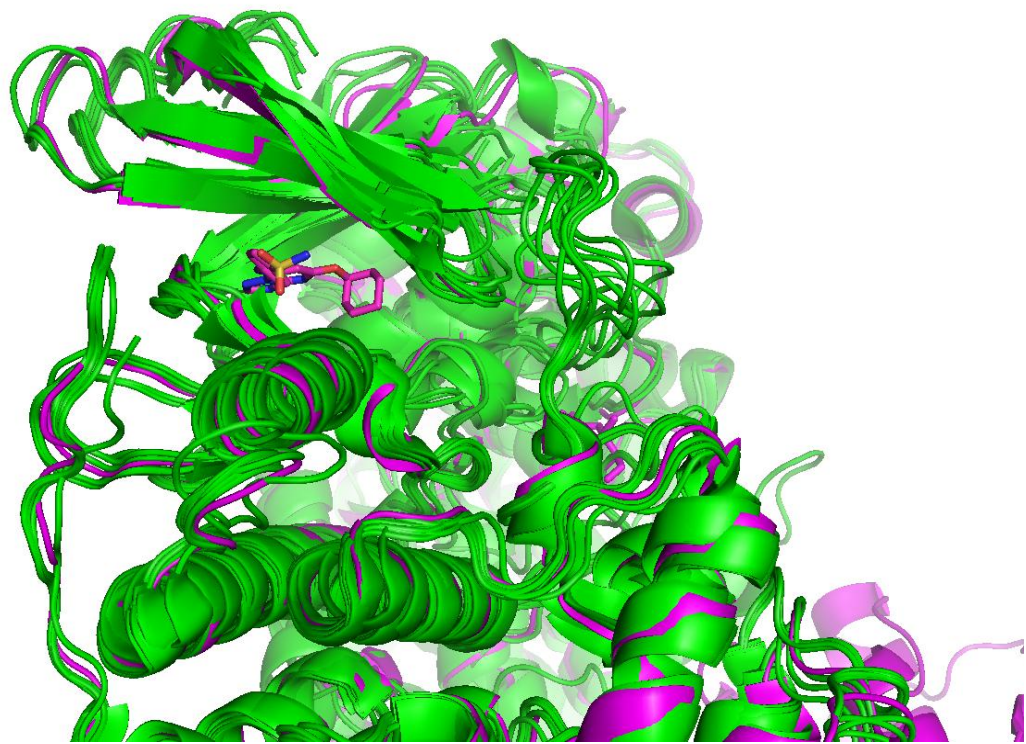

**Figure S2.** Histograms of the variations of distance between the binding site atoms of DHFR: A. CA-V8 and CZ- F34 ; B: CA-V8 and CA-F34; "0" represents the reference X-ray structure. Positive values represent an increase of the distance, and negative values represent a decrease of the distance.

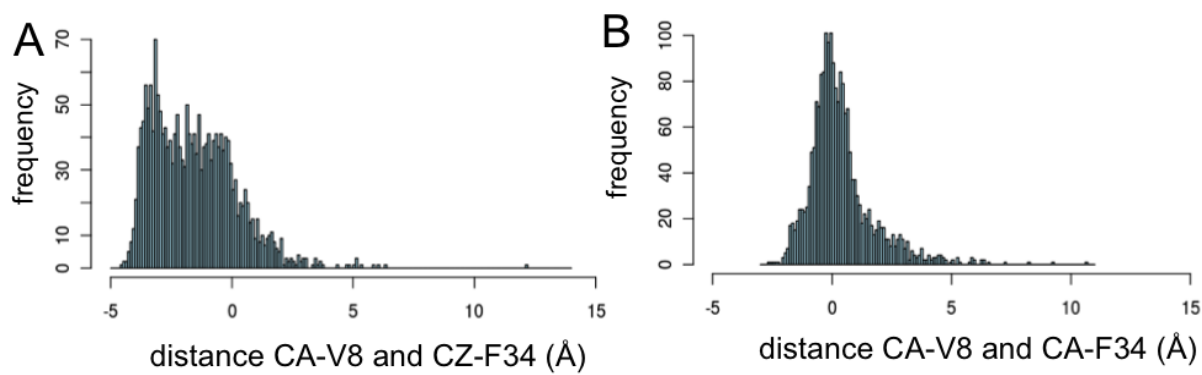

**Figure S3.** A histogram of the average of the binding energies calculated using Vina scoring for the interactions of the 2000 DHFR structures generated by the combined normal modes and the 191 known ligands.

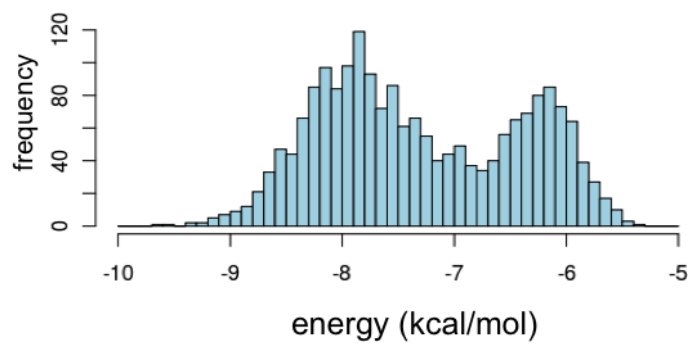

**Table S1.** Enrichment obtained on the five best performing NMA structures of DHFR having volumes of the binding site within a range of the volume of the X-ray active site ±30% at 1%, 5% and 10% of the screened chemical library. Volume and Drug score values of the active sites are calculated using DoGSite webserver [58].

| Structure | 1% | 5% | 10% | Volume | Drug Score |
|---|---|---|---|---|---|
| 35 1.8 | 1.6 | 16.2 | 28.3 | 340.29 | 0.62 |
| 18 1.4 | 2.6 | 15.2 | 27.2 | 302.85 | 0.54 |
| 24 -2.0 | 3.1 | 15.2 | 23.0 | 221.82 | 0.44 |
| 22 2.0 | 2.6 | 14.7 | 20.9 | 187.52 | 0.28 |
| 20 -1.4 | 3.7 | 13.6 | 21.5 | 389.03 | 0.66 |

**Table S2.** Enrichment on the five best performing structures of DHFR among the 8 centroids resulted from the Hclust and Kmeans clustering on the 2000 conformations generated from the combined normal modes at 1%, 5% and 10% of the screened chemical library. Volume and Drug score values of the active sites are calculated using DoGSite webserver [58].

| Structure | 1% | 5% | 10% | Volume | Drug Score |
|---|---|---|---|---|---|
| 926 | 1.0 | 9.9 | 15.7 | 241.2 | 0.44 |
| 940 | 1.0 | 11 | 14.7 | 469.7 | 0.45 |
| 1117 | 1.6 | 8.9 | 12 | 350.7 | 0.57 |
| 973 | 1.6 | 6.8 | 14.7 | 587.2 | 0.45 |
| 1053 | 1.0 | 7.9 | 14.1 | 912.8 | 0.56 |